\documentstyle[11pt,moriond,epsfig]{article}

\def\Journal#1#2#3#4{{#1} {\bf #2}, #3 (#4)}

\def\mnras{\em Mon. Not. R. astr. Soc.}
\def\aj{\em Astron. J.}
\def\apj{\em Astrophys. J. }
\def\aap{\em Astr. Astrophys. }

\def\aas{\em Astr. Astrophys.  Suppl.}

\def\nat{\em Nature }
\def\pasp{\em Pub.Astron.Soc.Pacific}

\def\be{\begin{equation}}
\def\ee{\end{equation}}
\def\bea{\begin{eqnarray}}
\def\eea{\end{eqnarray}}

\begin{document}
\vspace*{4cm}
\title{SYNCHROTRON EMISSION FROM THE GALAXY}

\author{ R.D. DAVIES and A. WILKINSON }

\address{Department of Physics \& Astronomy, Jodrell Bank, University
of Manchester\\
Manchester M13 9PL, England}

\maketitle\abstracts{Galactic synchrotron emission is a potentially
confusing foreground, both in total power and in polarization, to the
Cosmic Microwave Background Radiation. It also contains much physical
information in its own right. This review examines the
amplitude, angular power spectrum and frequency spectrum of the
synchrotron emission as derived from the presently available
de-striped maps. There are as yet no maps at arcminute
resolution at frequencies above 2.4 GHz.  This incomplete information
is supplemented with data from supernovae, which are thought to be the
progenitors of the loops and spurs found in the Galactic emission.
The possible variations of the frequency spectral index from pixel to pixel are
highlighted.  The relative contributions of free-free and synchrotron
radiation are compared, and it is concluded that the free-free
contribution may be smaller than had been predicted by COBE.  New high
resolution polarization surveys of the Galactic plane suggest detail
on all scales so far observed.  At high latitudes the large percentage
polarisation means that the foreground contamination of the polarised
CMB signal will be more serious than for the unpolarized
radiation.}

\section{Introduction}

	Galactic emission at radio wavelengths is important to
understand in its own right.  Moreover it is crucial to be able to
quantify and remove this component as a foreground to the cosmic
microwave background (CMB).  Both synchrotron and free-free emission
contribute to this foreground, with the synchrotron emission
dominating at low frequencies ($\leq$1 GHz).

	The synchrotron emissivity is a function of both the
relativistic (cosmic ray) density and the local magnetic field
strength.  The luminosity at frequency $\nu$ is given by 

\be
 I(\nu) = const \;  L N_0 B^{(p+1)/2} \nu^{-(p-1)/2}
\ee

\noindent where $N_0$ is the density of relativistic electrons, $L$ is the
emission depth, $B$ is the magnetic field and the relativistic
electron energy spectrum\cite{ve} is given by $dN/dE = N_0 E^{-p}$.  The
radio spectral index is $\alpha= (p-1)/2$ in energy terms or
$2+\alpha$ when expressed as a brightness temperature $T_B \propto
\nu^{-(2+\alpha)} = \nu^{-\beta}$.  Within the interstellar magnetic
field of 2 to 5 microgauss, emission at GHz frequencies is
characteristically from relativistic electrons with an energy of 1 to
10 GeV.  Both $B$ and $N_0$, as well as $p$, will vary from point to point in
the Galactic disk and nearby halo.  The cosmic ray electrons are
thought to originate mainly in supernovae then diffuse outwards in the
expanding remnant.  Structure will be formed in the remnant as it
collides with the non-uniform ambient medium.  The magnetic field will
be likewise amplified in compression regions and vary in strength and
direction.  The net effect is to produce elongated synchrotron
emission structures on a wide range of scales.  The spectral index of
the emission will vary with position for two reasons.  Firstly the
electron spectral index varies from one supernova to another and
secondly the spectrum steepens ($\Delta p = +1$) with time due to radiation
energy loss thus giving an age-dependent spectral index.

	This paper will describe the synchrotron features in and near
the Galactic plane which are believed to give rise to the structures
seen at higher galactic latitudes.  All-sky and large area surveys are
assessed to give information about the amplitude and spectrum of the
high latitude emission which is a potential confusing foreground to
the CMB.  Comments are given about the role of synchrotron
polarization and of free-free emission.

\section{Large area surveys at low frequencies}

	Radio surveys at frequencies less than 2 GHz are dominated by
synchrotron emission.  The well-known survey by Haslam et
al. \cite{ha} at 408 MHz is the only all-sky map available.
Large-area surveys with careful attention to baselines and calibration
have also been made at 1420 MHz \cite{rr1} and most recently at 2326
MHz \cite{jo}.  All these investigations have been made with FWHP
beamwidths of less than 1$^{\circ}$.  Before these surveys can be used
to derive the angular power spectrum and the emission (frequency)
spectral index, it is necessary to remove the baseline
stripes \cite{da} in the most commonly used radio maps at 408 and 1420
MHz.  These stripes contain power on angular scales of a few to ten
degrees.  Lasenby \cite{la} has used the 408 and 1420 MHz surveys to
estimate the spatial power spectrum of the high latitude region
surveyd by the Tenerife CMB experiments; he found an angular power
spectrum somewhat flatter than the $l^{-3}$ law derived for HI and for
IRAS far infrared emission.

	The spectral index of Galactic synchrotron emission can be
readily determined at frequencies less than 1 GHz where the
observational baselevel uncertainty is much less than the total
Galactic emission.  Lasenby \cite{la} used data covering the range
38 to 1420 MHz to determine the spectral index variation over the
northern sky.  Clear variations in spectral index of at least 0.3
about a mean value of 2.7 were found.  There was a steepening in the
spectral index at higher frequencies in the brighter features such as
the loops and some SNRs.  Up to 1420 MHz, no such steepening was found
in the regions of weaker emission.  At higher Galactic latitudes where
no reliable zero level is available at 1420 MHz, an estimate can be
made of the spectral index of local features by using the T-T
technique.  The de-striped 408 and 1420 MHz maps gave spectral indices
of $\beta$ = 2.8 to 3.2 in the northern galactic pole regions \cite{da}.

	The de-striped 408 MHz map shows there are substantial areas
($100^{\circ} \times 50^{\circ}$ in RA $\times$ Dec) in both the northern and
southern skies which are devoid of appreciable synchrotron structure
and can be used for CMB studies.  The Tenerife experiments have been
based on the northern low emissivity band centred on Dec =
40$^{\circ}$, RA = 130$^{\circ}$ to 250$^{\circ}$ where the rms
Galactic emission in a 5$^{\circ}$ beam is $\leq 4 \mu $K at 33 GHz
\cite{ha}.

\section{What we know from SNRs}

	Supernovae are probably the progenitors of the main structures
in the Galactic radio emission at intermediate latitudes.  The
supernova remnants (SNRs) are the easily recognisable early stages of
the expansion phenomenon.  A supernova releases about $10^{51}$ ergs into
the interstellar medium (ISM).  This passes through the free expansion
phase, then after it has encountered its own mass in the ISM it moves
into the Sedov (adiabatic) phase.  The SNR shock ultimately disappears
when the expansion velocity slows to the local sound velocity.  This
process takes $10^5$ to $10^6$ years to reach the point where the remnant is
not clearly recognizable as a single entity but will still give rise
to synchrotron emission from the residual CR electrons and magnetic
fields.

	An examination of statistics and structure of SNRs will give
some indication of the properties of the emission from their residual
structures.  Firstly, it is possible to recognise the various
evolutionary phases in the SNR phenomenon in individual remnants.
Objects like Cas A and the Crab Nebula (500 to 1000 years old) are in
the early free expansion stage while the Cygnus Loop (15 to 20,000
years old) is losing energy in multiple shocks.  Secondly, the
statistics of the spectral index of the integrated emission show a
spread of $\alpha$ of $\pm 0.1$ about the mean value of 0.7 \cite{rs},
with some objects such as the Crab Nebula ($\alpha$ = 0.3)
having more extreme indices.  This spread will presumably result in a
spread in the spectral indices of their residual structures.

	The Cygnus Loop provides an excellent case study of an SNR in
its late phase of evolution.  High sensitivity maps of arcmin
resolutions are available at a range of frequencies.  This remnant is
3.$^{\circ}$5 x 2.$^{\circ}$5 (30 pc $\times$ 20 pc) in diameter lying
at l = 74$^{\circ}$, b = -8.$^{\circ}$5 and at a distance of 500-800
pc.  The 1.4 GHz synthesis map by Leahy, Roger and Ballantyne
\cite{le} made with a 1x1 arcmin$^2$ beam shows both filamentary and
diffuse structure on angular scales from a few to 30 arcmins.  By
comparing maps at 0.408 and 2.695 GHz Green \cite{gr} finds
significant variations (0.3) in spectral index between the major
features of the remnant.

\section{What we know from spurs and loops}

	Large features with a synchrotron spectrum extend far from the
Galactic plane.  The most prominent of these are the spurs and loops
which describe small circles on the sky with diameters in the range
60$^{\circ}$ to 120$^{\circ}$ \cite{be}.  Because of their association
with HI and, in some cases, with X-ray emission, they are believed to
be low surface brightness counterparts of the brighter SNRs seen at
lower latitudes.  Other more diffuse structure at higher latitudes may
be even older remnants.  Reich and Reich \cite{rr2} find that the
Loops I (the North Polar Spur) and II have a steeper spectrum than the
average Galactic emission.  Between 408 MHz and 1420 MHz these two
loops have temperature spectral indices of 2.93 and 2.85 respectively.
Using T-T plots Davies, Watson and Gutierrez \cite{da} derived a
spectral index of $\beta$ = 3.2 for the brightest part of Loop I.
Lawson et al.\cite{la} claim that most of the large structures seen
on the maps are related to Loops I and III suggesting that they are
the evidence of diffusive shock acceleration of the CR electrons
derived from the supernova.  They derive a distance for Loop I of 130
$\pm$ 75 pc and a radius of 115 $\pm$ 68 pc; Loop III is thought to be
of similar size.

	A revealing high sensitivity survey of a substantial part of
the southern Galactic plane (l = 238$^{\circ}$ to 365$^{\circ}$, b =
-5$^{\circ}$ to +5$^{\circ}$) has been made with a resolution of 10
arcmin at 2.4 GHz by Duncan et al. \cite{du1}.  These authors found a
large amount of structure and detail including many low surface
brightness loops and spurs.  They also list over 30 possible SNR
candidates, a number of which have angular diameters of about
10$^{\circ}$.  Many of the spurs can be traced even further from the
plane in the 2.326 GHz survey of Jonas, Baarth and Nicolson
\cite{jo}. The spectral index of these new spurs has still to be
determined.

\section{Higher frequency surveys at Jodrell Bank and Tenerife}

	It is important to make Galactic surveys at intermediate and
high latitudes at frequencies closer to those at which the CMB
structure is being investigated.  Because of the variation of spectral
index from one region to another, the structure at 408 MHz will be
quite different from that at 10 or 30 GHz.  Two separate
experiments have been established to address this problem.  The first
is a high sensitivity short baseline interferometer operating at 5 GHz
at Jodrell Bank \cite{me}(Fig. 1).  By making observations at a
range of baselines the point source and the Galactic contribution to
the microwave background can be separated.  After correction for the
point sources the Galactic emission at 5 GHz can be compared with the
published 408 MHz survey.  The spectral index of Galactic features in
the survey was found to be about 3.0 at intermediate and higher
Galactic latitudes.

	The 10, 15 and 33 GHz beamswitching radiometers at Teide
Observatory Tenerife are scaled to give the same resolution
(5$^{\circ}$ FWHP beamswitched $\pm$8$^{\circ}$).  Comparison of the
surveys at the three frequencies should allow the Galactic emission to
be separated from the intrinsic CMB component.  Observations already
available from this experiment can constrain the Galactic contribution
at 33 GHz.  The observed rms fluctuation level at 10 GHz in the Dec =
+40$^{\circ}$ scan between RA = 160$^{\circ}$ and 230$^{\circ}$ is 29
+20/-30 $\mu$K \cite{ha}.  This rms level will include both a Galactic
and CMB contribution.  Using a 2$\sigma$ upper limit and subtracting
quadratically the detected CMB signal (54 $\mu$K) we derive an upper
limit for the Galactic emission at 10 GHz of 43 $\mu$K.  This would
produce an upper limit at 33 GHz of 2 $\mu$K if it were synchrotron
with $\beta$ = 3.0 and an upper limit of 4 $\mu$K if it were free-free
emission with $\beta$ = 2.1.  Hancock et al. \cite{ha} derive a
spectral index for Galactic emission at the 3 frequencies of $\beta$ =
3.1.

\section{ Synchrotron versus free-free}

	Free-free emission is not easily identified at radio
frequencies except near the Galactic plane.  At higher latitudes it
must be separated from synchrotron emission by virtue of its different
spectral index.  At higher frequencies where free-free emission might
be expected to exceed the synchrotron component, the signals are weak
and the survey zero levels are indeterminate.  Most of the information
on the thermal electron content currently available at intermediate
and higher frequencies comes from H$\alpha$ surveys.  This diffuse
H$\alpha$ emission is thought to be a good tracer of diffuse free-free
emission since both are emitted by the same ionized medium and both
intensities are proportional to emission measure ($EM = ne^2 dl$), the
line of sight integral of the free electron density squared.  Major
H$\alpha$ structures are a feature of the well-known Local (Gould
Belt) System which extends some 40$^{\circ}$ from the plane at
positive $b$ in the Galactic centre and at negative $b$ in the
anticentre.  Other H$\alpha$ features are known to extend 15$^{\circ}$
to 20$^{\circ}$ from the plane \cite{si}.

\begin{figure}
\epsfig{file=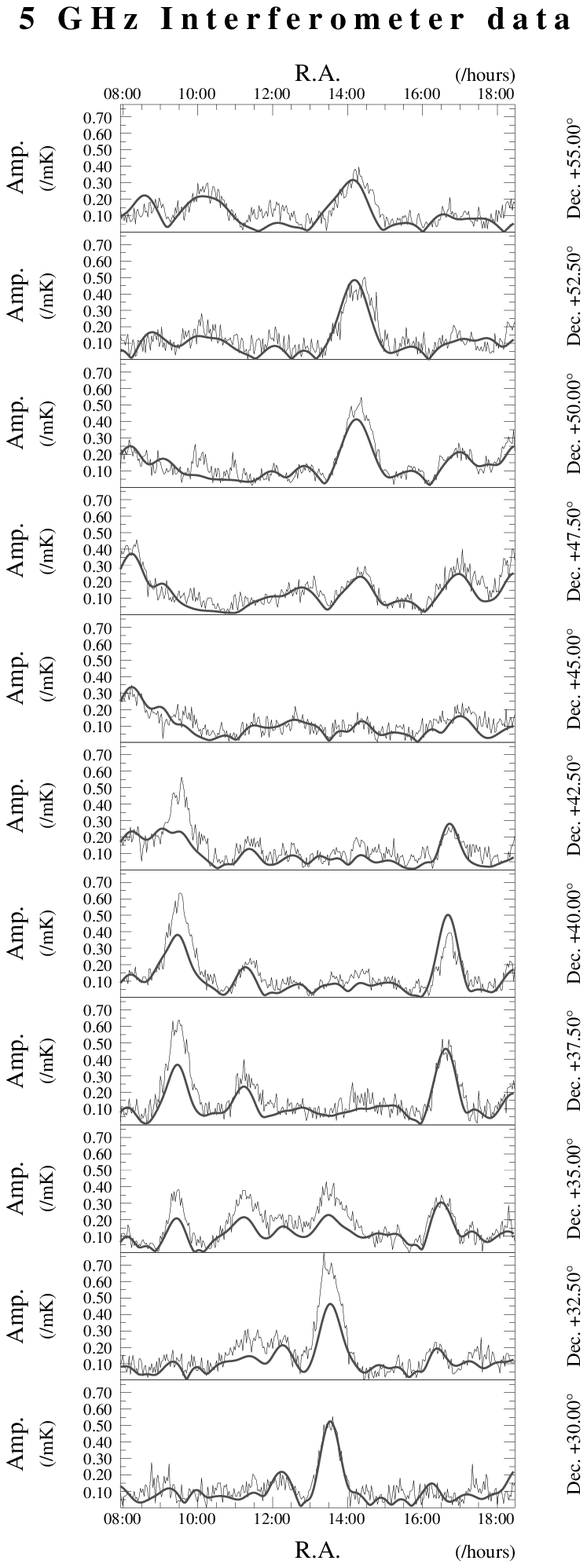}
\caption{Comparison of 5 GHz interferometer narrow spacing (baseline 12
wavelengths ) data with the Greenbank point source survey (dark line).  The
differences are due to source variability or Galactic synchrotron emission.}
\end{figure}
	The intermediate latitude H$\alpha$ distribution may be modelled
(Reynolds 1992) as a layer parallel to the Galactic plane with a
half-thickness intensity of 1.2 Rayleigh (R).  The rms variation in
this H$\alpha$ emission is about 0.6R on degree scales.  In the context of
the present discussion 1R will give a brightness temperature of about
10 $\mu$K at 45 GHz.  Further information about angular structure in the
H$\alpha$ emission can be derived from the North Celestial Pole (NCP) study
by Gaustad et al.\cite{ga}.  Veeraraghavan \& Davies \cite{ve} used this
material to derive a spatial power spectrum on angular scales of 10
arcmin to a few degrees.  The spatial power law index is -2.3 $\pm$ 0.1
over this range with an rms amplitude of $0.12 cosec(b)$ Rayleighs on 10
arcmin scales.  This level is consistent with the limits derived from
the Tenerife experiments and indicates that the free-free rms
brightness becomes comparable with the synchrotron value at about 20
GHz where it would comprise 10 to 20 percent of the CMB fluctuation
amplitude.  Kogut et al. \cite{ko} compared the COBE DMR maps with the
DIRBE maps and claimed a correlation between free-free and dust
emission.  They obtained free-free levels somewhat higher (in a 7$^{\circ}$
beam) than measured more directly in the Tenerife experiments.

\subsection{	A note on polarization}

	Extensive surveys of the polarization of Galactic synchrotron
polarization have been made at frequencies up to 1-2 GHz.  Significant
polarized signals are found over most of the surveyed sky.  The
percentage polarization increases with frequency indicating the
presence of a Faraday rotating medium with a rotation measure of ~8
rad m$^{-2}$ \cite{sp}.  The mean polarization amplitude at 1400 MHz
lies between 20 and 30 \% at higher ($|b| \geq 30^{\circ}$) Galactic latitudes
as seen in an 0.$^{\circ}$6 beam.  The polarization degree in Loop I reaches
$\sim$72 \% at higher Galactic latitudes.  This is close to the
theoretical upper limit where the fractional polarization is given by
\be
		\pi  =  \frac{3\beta-3}{3\beta-1}
\ee 		
where $\beta$ is the temperature spectral index.
The other loops have maximum polarization in the range 30-50 \%.

	A detailed study of the polarization of the Cygnus Loop has
been made at 1.4 GHz by Leahy, Roger \& Ballantyne \cite{le}.  They find
that the bright filaments have the B field aligned along their length
with a maximum polarization in the remnant of 39 percent and a mean
value of 7 percent.  The lower values of polarization in some areas
are most likely due to depolarization in the Faraday screen of the
object which has a rotation measure of -20 to -35 rad m$^{-2}$.  The 5 GHz
map of Kundu \& Becker \cite{ku} shows a fractional polarization of 25
percent over the southern half of the source.

	The 2.4 GHz Galactic plane survey by Duncan et al. \cite{du2}
shows considerable complex structure with their 10 arcmin beam.
Bright, extended regions of polarization emission of the order of 5$^{\circ}$
across include the Vela SNR and a large structure appearing to the
north of Sgr A.  A quasi-uniform weak component of patchy polarization
is seen over the length of the survey.

	Theory indicates that the CMB radiation will be polarized at a
level of 5 to 10 percent.  On the other hand the synchrotron emission
can be 30 percent polarized at 1.4 GHz and probably higher at higher
frequencies.  Accordingly this foreground Galactic polarization must
be considered more seriously than the total power case when measuring
CMB polarization.  A foreground feature which is 10 percent of an
intrinsic CMB feature may have polarization which is 30 to 50 percent
of the polarized intensity of the feature.

\section*{Acknowledgments}
AW would like to acknowledge the receipt of a Daphne Jackson Research
Fellowship, sponsored by PPARC.

\end{document}